\documentclass[aps,pre,twocolumn,10pt]{revtex4-1}

\usepackage{graphicx}
\usepackage{amssymb}
\usepackage{amsmath}

\renewcommand{\eqref}[1]{Eq. (\ref{#1})}
\newcommand{\bms}[1]{\boldsymbol{#1}}
\newcommand{\arho}{\bar{\rho}}
\newcommand{\nn}{\nonumber}

\begin{document}

\title{Vapor--liquid coexistence of the Stockmayer fluid
in nonuniform external fields} %
\author{Sela Samin} %
\author{Yoav Tsori}

\affiliation{Department of Chemical Engineering and The Ilse Katz Institute
for
Nanoscale
Science and Technology, Ben-Gurion University of the Negev, 84105 Beer-Sheva,
Israel.}

\author{Christian Holm}

\affiliation{Institut f\"{u}r Computerphysik, Universit\"{a}t Stuttgart,
Allmandring 3, D-70569 Stuttgart,Germany}

\begin{abstract}

We investigate the structure and phase behavior of the Stockmayer fluid in the
presence of nonuniform electric fields using molecular simulation. We find that
an initially homogeneous vapor phase undergoes a local phase
separation in a nonuniform field due to the combined effect of the field
gradient and the fluid vapor--liquid equilibrium. This results in a high density
fluid condensing in the strong field region. The system
polarization exhibits a strong field dependence due to the fluid condensation.

\end{abstract}

\maketitle

\section{Introduction}

The application of external fields for manipulation of the structure and phase
behavior of dipolar fluids has attracted growing interest in recent years
\cite{klapp_review,holm2005}. Typically, both theoretical and experimental
studies consider uniform applied
fields. However, in complex systems such as microfluidic devices field gradients
occur naturally. 
Motivated by experimental work on the demixing of binary mixtures in field
gradients \cite{efdemix}, we have previously studied theoretically the
application of nonuniform electric fields to pure fluids \cite{efips_jpcb} and
simple mixtures \cite{efips_jcp1,efips_jcp2}. Another experimental realization
of the ability of field gradients to promote
phase transitions is the field induced crystallization of colloidal
suspensions promoted by dielectrophoretic forces
\cite{chaikin_prl2006,chaikin_jcp2008,lumsdon2004}.

In particular, we examined the
effect of nonuniform fields on the vapor--liquid coexistence \cite{efips_jpcb}
by combining the simple van der Waals mean field theory with Onsager's theory of
dielectrics to investigate polar and nonpolar fluids. Our main finding was that
above a critical field, in situations where the fluid is unperturbed by
a uniform field, a nucleation of a gas bubble from the liquid phase or a liquid
droplet from the vapor phase is induced by a nonuniform field. This phase
separation transition is promoted by the dielectrophoretic force which favors a
higher permittivity (density) fluid in the region of strong field. The resulting
modification in the fluid phase diagram is considerably larger compared to
uniform fields.

A system which can be considered as an idealized manifestation of
nonuniform fields is a grand canonical ensemble where a uniform field,
$\bms{E}=const.$, is applied within the system volume but not in the
material reservoir, where $\bms{E}=0$. An example is a slit pore in
equilibrium with a bulk fluid. Clearly, in real systems there exists
an interfacial region at the pore edges where the field is
nonuniform. The field effect in this type of system has been studied
for one component fluids \cite{bratko2007,bratko2008} where it was
found to gradually increase the fluid density within the pore
moderately. A stronger field effect was found by Brunet {\it et al.}
\cite{brunet2009,brunet2010} who studied mixtures. They observed that
if the mixture has a demixing instability and one of the components is
dipolar, the coupling to the external field leads to a pore filling
transition which allows a sensitive control of the pore composition.

The goal of this paper is to study the structure of a dipolar vapor
confined in a slit pore and exposed to nonuniform electric fields in
the canonical ensemble via molecular simulation. We compare our
results with mean field theory and discuss the consequences of the
full description of the vapor--liquid interface in finite systems. The
paper is organized as follows: Section \ref{sec:methods} discusses the
simulation methods and model system. Section \ref{sec:uniform}
compares results for the fluid in a uniform field with previous
studies. The results for a nonuniform field are compiled in Section
\ref{sec:nonuniform}. Conclusions are given in Section
\ref{sec:conc}.

\section{Simulation Methods}
\label{sec:methods}

We consider $N$ spherical dipolar particles with a diameter $\sigma$ and
a permanent dipole moment $\bms{\mu}$. The particles interact via the Stockmayer
pair potential:
\begin{align}
 U_{ij}&=4\epsilon\left[\left(\frac { \sigma } { \left|\bms{r}_{ij}\right| }
\right)^ { 12 }
-\left(\frac{\sigma}{\left|\bms{r}_{ij}\right|}\right)^6 \right] \nn \\
&+\frac{1}{4\pi\varepsilon_0}\left(\frac{\bms{\mu}_i\cdot\bms{\mu}_j}{\left|\bms
{ r } _ { ij} \right|^3 }
-\frac{(\bms{\mu}_i\cdot\bms{r}_{ij})(\bms{\mu}_j\cdot\bms{r}_{ij})}{
\left|\bms{r}_{ij}\right|^5}\right)
\end{align}
where $\bms{r}_{ij}=\bms{r}_{i}-\bms{r}_{j}$ stands for the
displacement vector of particles $i$ and $j$, $\epsilon$ denotes the
Lennard-Jones (LJ) interaction parameter and $\varepsilon_0$ is the
vacuum permittivity. In the following we use reduced units; length:
$r^*=r/\sigma$, temperature: $T^*=k_BT/\epsilon$, density:
$\rho^*=\rho\sigma^3$, dipole moment:
$\mu^*=\mu/\sqrt{4\pi\varepsilon_0\epsilon\sigma^3}$ and external
field: $E^*=E\sqrt{4\pi\varepsilon_0\epsilon\sigma^3}$. Here, $k_B$ is
the Boltzmann constant and we take
$\sigma=\epsilon=4\pi\varepsilon_0=1$. For brevity we omit the
asterisk superscript henceforth.

The phase diagram of bulk and confined Stockmayer fluids in
a uniform field was determined using a Gibbs Ensemble -- Hybrid Monte Carlo
scheme (GE-HMC). The classical Gibbs Ensemble Monte
Carlo (GEMC) simulation \cite{gibbs_ensemble} offers a simple 
method to determine the vapor--liquid equilibrium densities and pressures with
a single simulation run. This is achieved by equating the chemical potentials
and pressures of two simulation boxes using appropriate Monte Carlo particle and
volume exchange moves, respectively \cite{frenkel_book}. The third type of moves
performed are
particle translations, rotations and other conformational changes of
single particles within the two boxes. In the GE-HMC variation, single
particle moves are replaced by a collective Hybrid
Monte Carlo (HMC) move \cite{Duane1987,Mehlig1992}. A single HMC cycle consists
of three steps: first, particles of the current configuration, $o$, are assigned
new momenta and angular velocities by sampling a Gaussian
distribution corresponding to the desired temperature. Second, the new
configuration, $n$, is generated from a short MD trajectory in the
microcanonial ensemble. Lastly, the new configuration is accepted/rejected
according to the Metropolis criterion:
\begin{align}
 \min\left(1,\exp\left(-\Delta H/T \right)\right)
\end{align}
where $\Delta H=H(n)-H(o)$ is the resulting change in the system
Hamiltonian. Detailed balance is satisfied if the integration
algorithm used for the MD trajectory is time-reversible and
area-preserving \cite{Mehlig1992}, which is fulfilled by a simple
velocity-Verlet
integrator. All MD trajectories were produced with the ESPResSo
package \cite{espresso}. The collective HMC moves allow to efficiently sample
the
high density liquid phase and complex molecular configurations
\cite{desgranges2009}.

GE-HMC simulation cycles were conducted with 512 Stockmayer
particles. The total simulation consisted of $10^4$ cycles and
observable sampling was done after $2500$ equilibration cycles.
A single cycle was composed
of 100 MC moves where the
probability of the move type was as follows: 0.8 for a particle exchange move,
0.15 for a HMC move and 0.05 for a volume exchange move. The number of time
steps in a HMC move was 10. Both the time step and the attempted volume change
were adjusted during equilibration such that approximately $50\%$ of the moves
were accepted. 

\begin{figure}[!b]
\includegraphics[width=1.75in]{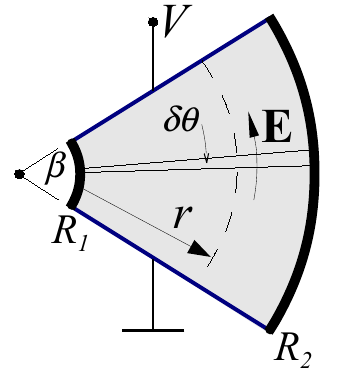}
\caption{The model condenser: a wedge made of two flat electrodes with a
potential difference $V$ across
them and an
angle $\beta$ between them. $R_1$ ($R_2$) is the distance of the inner
(outer) insulating wall from the imaginary meeting point of the electrodes. 
$r$ is the radial distance from the
inner wall (thick line). The resulting electric field $\mathbf{E}$ is along the
$\hat{\mathbf{\theta}}$ direction. In the simulations we approximate a small
angular segment $\delta \theta$ as a slit.}
\label{fig_system}
\end{figure}

Simulations in a nonuniform field where performed in a simplified model system,
see Fig. \ref{fig_system}.
Consider the fluid confined in a wedge condenser made up from two flat
electrodes with a potential difference $V$ across them and an angle $\beta$
between them. In this geometry the field
in the azimuthal direction is perpendicular to the field
gradient in the radial direction. This simplifies the solution of Gauss' law
since it implies $\nabla \varepsilon \cdot \mathbf{E} =0$ \cite{efips_jpcb}. The
resulting electric field is:
\begin{align}
\label{eq_wedge1}
\mathbf{E}(r)=\frac{V/\beta}{R_1+r} \hat{\mathbf{\theta}}
\end{align}
where $R_1$ is the radius of the inner condenser wall, $R_1+r$ the radial
distance from the imaginary meeting point of the electrodes and $\theta$ is the
azimuthal angle. We focus on a small angular section $\delta \theta$ far from
the electrodes of the capacitor and therefore rewrite \eqref{eq_wedge1} in
terms of Cartesian coordinates:
\begin{equation}
\label{eq_wedge2}
 \mathbf{E}(z)=\frac{E_0}{1+A_0z}\hat{x}
\end{equation}
where $E_0=V/(\beta R_1)$ is the maximal field at $z=0$ and
$A_0=1/R_1$ is a constant
characterizing the length scale of the field gradient; for $A_0=0$ the
field is uniform. 

In this paper we study the Stockmayer fluid under the external
field given in \eqref{eq_wedge2}. The contribution of this field to the 
potential energy of a single particle is given by
\begin{equation}
\label{eq_upot}
 U_{i}^{ext}=- \bms{\mu}_i \cdot
\bms{E}_i=-\frac{\mu_{i,x} E_0}{1+A_0z_i}
\end{equation}
where $\bms{E}_i$ is the field at the site of
particle $i$ which is a function of the particle coordinate $z_i$. Via the
potential energy we derive the additional force and torque on particle $i$ due
to
the field, i.e,
\begin{align}
\label{eq_force}
\bms{F}_{i}^{ext}&=-\frac{\mu_{i,x} A_0 E_0}{(1+A_0z_i)^2}\hat{z}\\
\label{eq_torque}
\bms{T}_{i}^{ext}&=\frac{\mu_{i,z} E_0}{1+A_0z_i}\hat{y}-\frac{\mu_{i,y}
E_0}{1+A_0z_i}\hat{z}
\end{align}
Note that the force contribution, \eqref{eq_force}, vanishes in the
case of a uniform field. Moreover, \eqref{eq_force} indicates that
particles are drawn to the strong field region and feel a stronger
force when aligned with the field. This may be thought of as the
microscopic origin of the dielectrophoretic force.

MD simulations of the Stockmayer fluid in a nonuniform field were
performed using the suitably modified ESPREesSo package
\cite{espresso}. During the simulation the LJ potential was cut off at
$r_{c}=3$ and the long range dipolar potential was evaluated using the
dipolar P$^3$M algorithm \cite{cerda2008} with metallic boundary
conditions.

When a nonuniform field is applied to the fluid its translational
invariance in the direction of the field gradient is broken and
therefore periodic boundary conditions (PBC) can not be used in this
direction. However, the implementation of the P$^3$M method requires
that we employ PBC in all directions. Therefore, in the simulations we
model the condenser as an infinite slab with the two confining walls
placed at $z=0$ and $z=D<L$, where $L$ is the cubic simulation box
length. The unwanted dipolar interactions between slabs replicated
along the $z$ direction are corrected using the dipolar layer
correction method of Ref. \cite{brodka2004}. This allows us to use a
small gap of empty space in the simulation box. In order to isolate
the field effect, we use for the fluid-wall interaction a purely
repulsive LJ potential shifted and cut off at $r_{c}=2^{1/6}$ (WCA
potential).

Simulations of the dipolar fluid in the slab where initialized from
random particle configurations. In the simulations we employ a
Langevin thermostat. A time step of
$\Delta t = 0.004$ was used in all simulations. The simulations
equilibration period varied from $1\times10^5-6\times10^5$ time steps,
depending on the field strength. The structural and dielectric
properties of the system were then sampled every 200 time steps for at
least $4\times10^5$ time steps. Time averaged quantities sampled are
denoted by $\langle ... \rangle$.

\section{Results and Discussion}

\subsection{Phase behavior of the Stockmayer fluid in a uniform field}
\label{sec:uniform}

\begin{figure}[!tb] 
\begin{center}
 \includegraphics[width=3.3in]{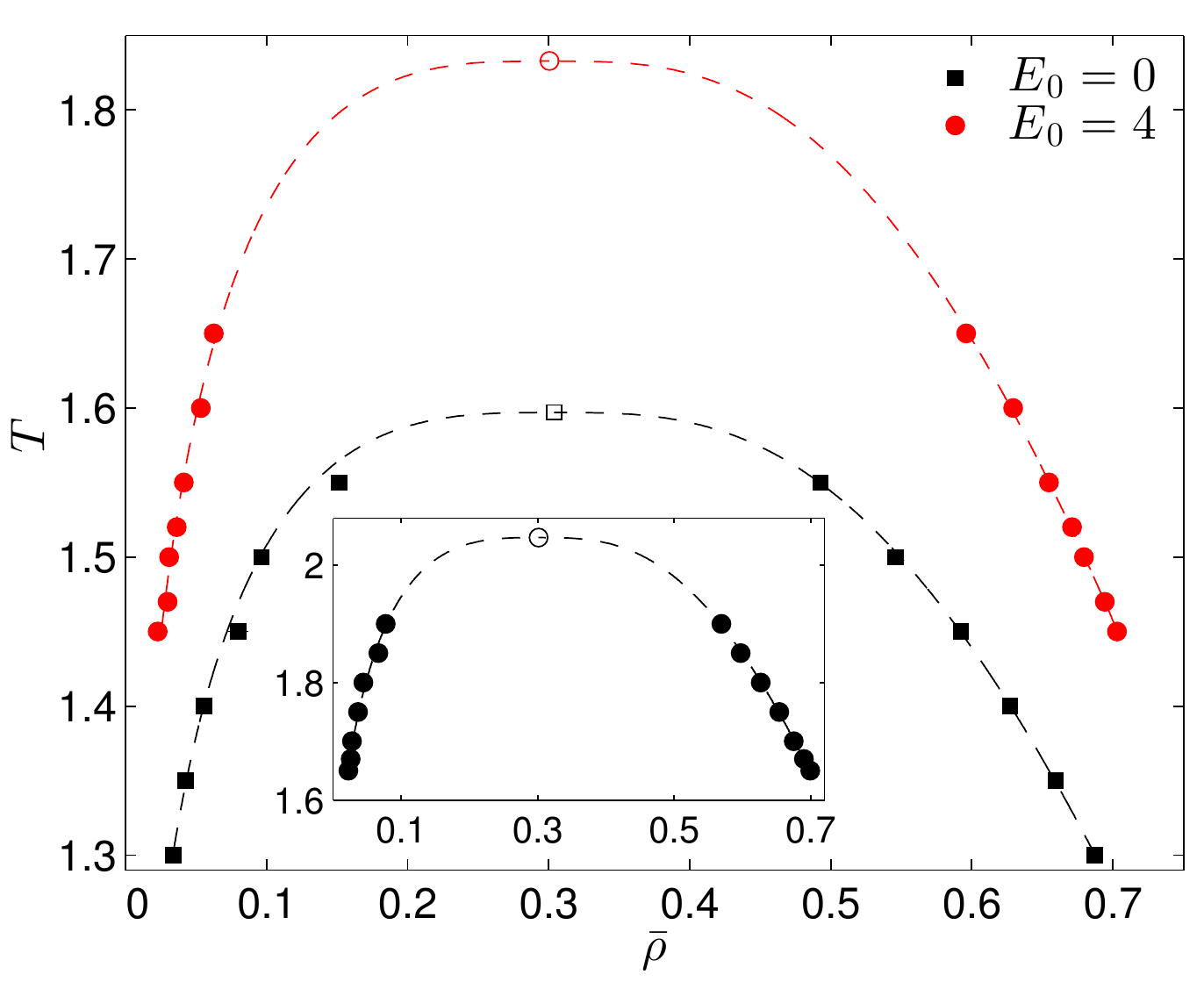}
 \caption{Vapor--liquid coexistence of the bulk Stockmayer fluid with
  $\mu=1.5$. Squares: in the absence of an external field. Circles:
  with a uniform field $E_0=4$. Inset: in the absence of an external
  field for $\mu=2$. Dashed curves are fits to the law of
  rectilinear diameters using the Ising exponent
  $\beta=0.326$. Critical points are marked by hollow symbols.}
\label{fig_pd}
\end{center} 
\end{figure}

We first tested the applicability of our GE-HMC simulation by
comparing our results to available data on the Stockmayer fluid with
$\mu=2$. Here, the standard long range correction is applied to the
LJ interaction \cite{frenkel_book}. The resulting coexistence curve is
shown in the inset of Fig. \ref{fig_pd}. We obtain a critical
temperature $T_c=2.05$ and density $\rho_c=0.301$, in good agreement
with the results of Van Leeuwen {\it et al.} \cite{leeuwen1993}
($͑T_c=2.06$, $\rho_c=0.289$) and Kiyohara {\it et al.}
\cite{kiyohara1997} ($T_c=2.05$, $\rho_c=0.306$). The small
differences in critical parameters are probably due to fact that
unlike the works above, the LJ potential in this study is cut off at a
fixed radius.

Henceforth, we will focus in this work on the Stockmayer fluid with
$\mu=1.5$. Vapor--liquid coexistence curves for such a bulk
Stockmayer fluid with and without an external field are shown in Fig.
\ref{fig_pd}. Here, no long range correction is applied to the LJ
interaction since we intend to compare these results with those
obtained in the slab geometry. In the absence of an external field we
find $T_c=1.59$ and $\rho_c=0.305$. In accord with previous studies,
we find when a uniform field is applied the unstable region in the
phase plane is increased
\cite{stevens1995,boda1996,kiyohara1999,szalai2003,jia2009}. This is
due to the increased dipole-dipole interaction and correlation as the
dipoles get aligned by the field \cite{stevens1995,jia2009}. In
particular, simulations of the Stockmayer fluid in an external field
found reasonable agreement with the Landau mean field theory
\cite{landau} for the field effect on the critical temperature
\cite{jia2009}. For $E_0=4$ we find that the critical temperature
increases to $T_c=1.83$ and $\rho_c=0.301$.

The value of $\mu=1.5$ for the dipole moment was chosen since it is
suited for description of both molecular fluids \cite{vanleeuwen1994}
and dipolar colloidal suspension alike \cite{holm2005}. Furthermore,
we will from now on set $T=1.6<T_c$. Hence, the value of dipolar
coupling constant, $\lambda=\mu^2/ T$, is $\lambda=1.41$. This means
that the dipolar interaction at a distance $\sigma$ is comparable to
both the thermal and LJ interactions. Note that a $\lambda$ value
close to unity below the critical temperature can only be realized for
relatively small values of $\mu$ \cite{bartke2007}.

\begin{figure}[!tb] 
\begin{center}
 \includegraphics[width=3.3in]{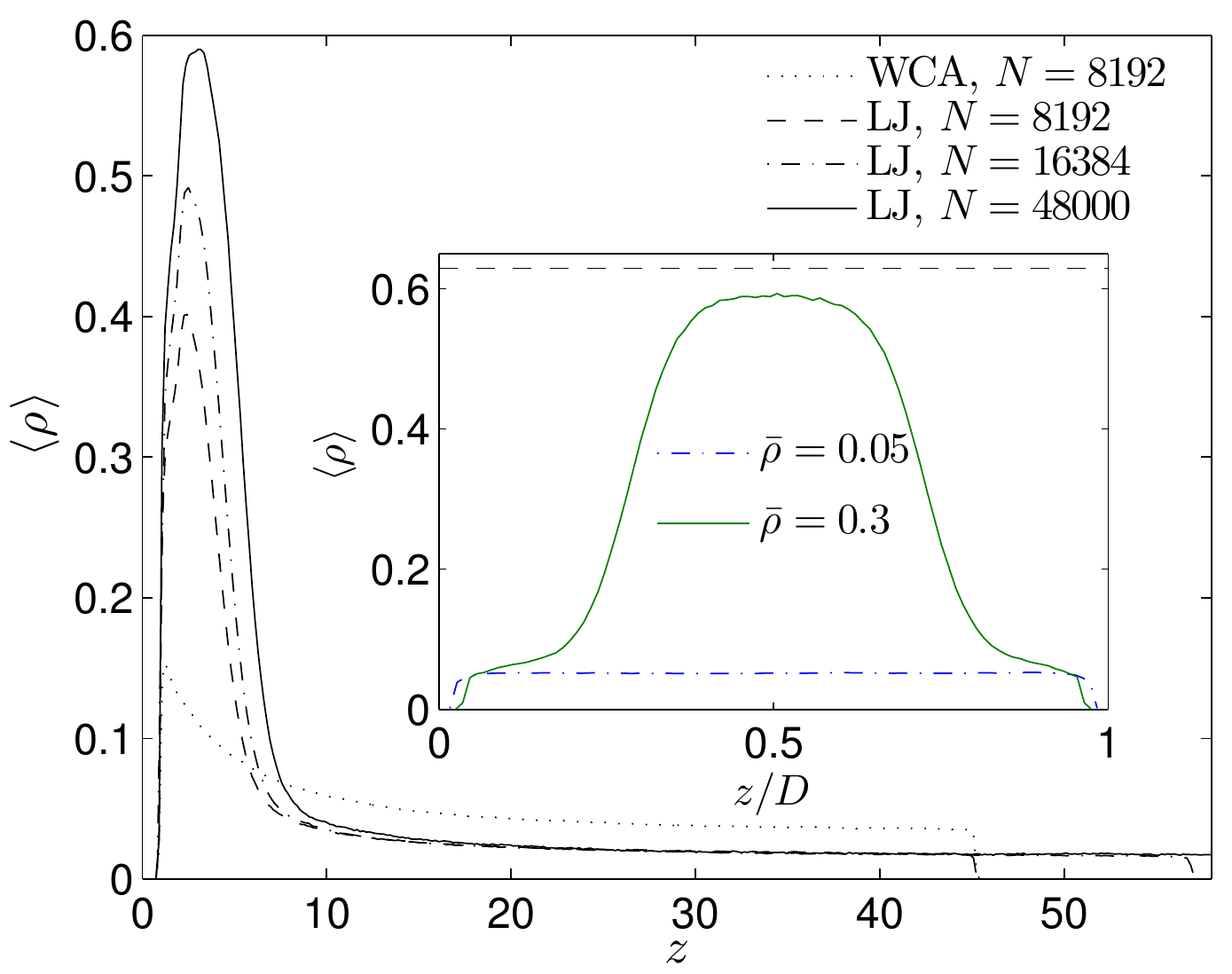}
 \caption{Density profiles $\langle \rho(z) \rangle$ in a nonuniform
  field for a Stockmayer fluid with an average density $\arho=0.05$
  and a temperature $T=1.6$. For the field we used a magnitude of
  $E_0=4$ in \eqref{eq_wedge2}. The solid curve shows results for
  $N=8192$ particles using $A_0=0.1$ in \eqref{eq_wedge2}. For the
  dash-dot curve $N=16384$ and $A_0=0.0794$; for the dashed curve
  $N=48000$ and $A_0=0.0555$. Dotted curve: same as the dashed curve
  but replacing the LJ part of the potential by a WCA
  potential. Inset: density profiles in a uniform field
  $\bms{E}=4\hat{x}$ at $T=1.6$. Two average densities, $\arho=0.3$
  and $0.05$ are presented corresponding to slab widths of $D\approx
  25$ and $D\approx 46$, respectively.}
\label{fig_compare}
\end{center}
\end{figure}

The effect of confinement on the coexistence curve of the Stockmayer fluid
is that of suppression of the unstable region of the phase diagram. Studies
conducted so far focussed on narrow slabs where this effect is large
\cite{richardi2008}. The finite-size effects for wider slabs where not
studied since they are quite small and also computationally prohibitively
expensive. 

An estimate of finite-size effects is provided in the inset of Fig.
\ref{fig_compare}, which
shows the density profiles, $\langle \rho(z) \rangle$,
for $N=8192$ confined particles at $T=1.6$ and under an uniform field
$\bms{E}=4\hat{x}$, parallel to the slab walls. For an average fluid density
$\arho=0.3$ (slab width
of $D\approx 25$), close to the critical density, the density profile exhibits
vapor--liquid coexistence. The dashed horizontal line in the inset corresponds
to the bulk liquid density of $\rho_l=0.629$ which is only slightly higher than
the average density of $\approx 0.59$ for the liquid in the slab. In contrast,
slowly expanding the slab such that a fluid average density of $\arho=0.05$ is
finally obtained (slab width of $L\approx 46$) results in a homogeneous vapor
phase, see the dash-dot curve in the inset of Fig. \ref{fig_compare}. This is
expected since $\arho=0.05$ is smaller than the vapor phase density of the bulk
fluid $\rho_v=0.053$.

\subsection{The Stockmayer fluid in a nonuniform field} \label{sec:nonuniform}

\begin{figure}[!tb] 
\begin{center}
 \includegraphics[width=3.3in]{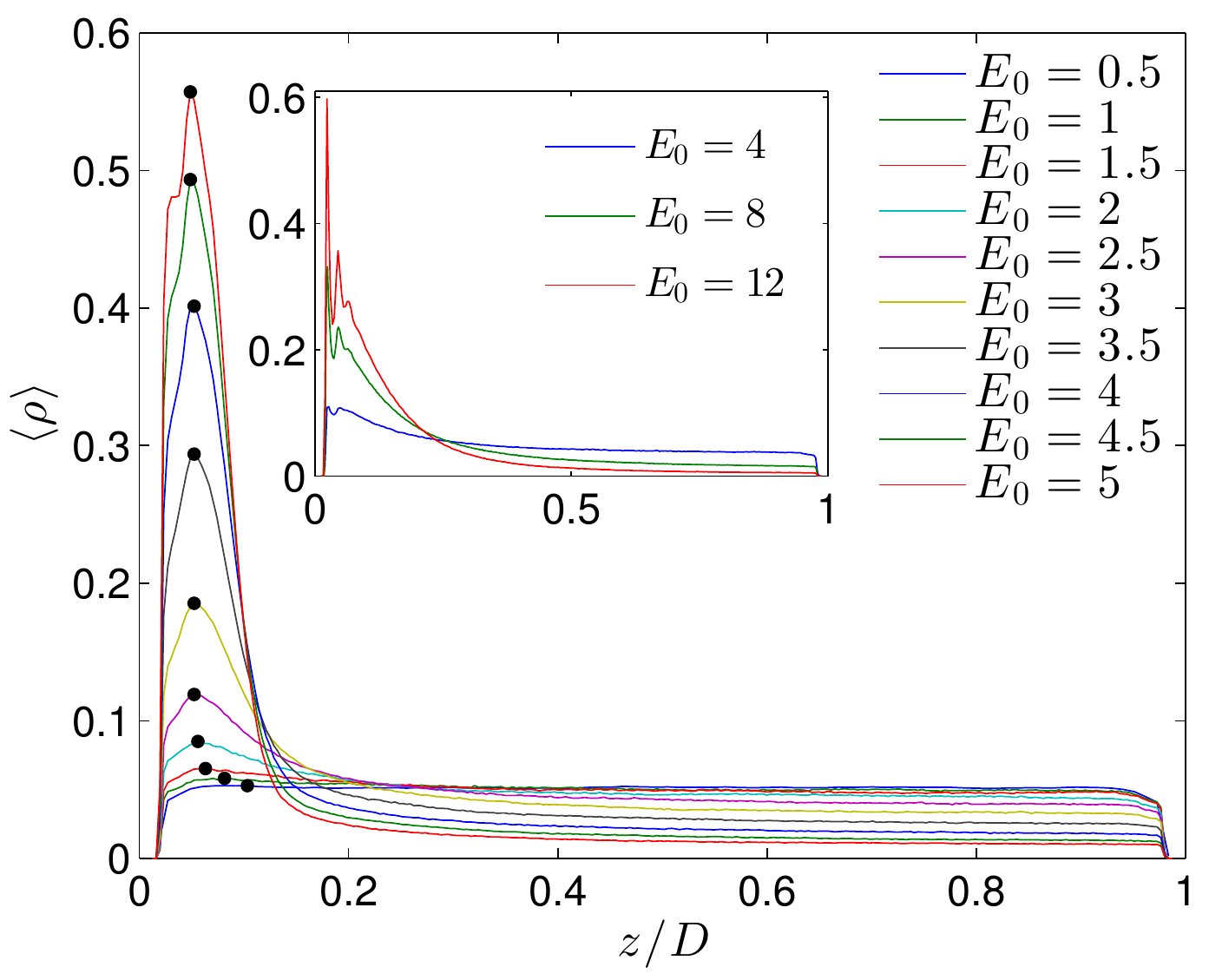}
 \caption{Density profiles $\langle \rho(z) \rangle$ in a nonuniform
  field of different magnitudes. The fluid maximal density is in the
  strong field region close to the wall (circle markers) and it
  increases with the field magnitude. Inset: density profiles for a
  nonuniform field perpendicular to the the slab walls (see text).
  For all curves $A_0=0.1$, the fluid average density is $\arho=0.05$
  and the temperature is $T=1.6$.}
\label{fig_nuf1}
\end{center} 
\end{figure}

The situation is markedly different when a nonuniform field is applied. The
dashed curve in Fig. \ref{fig_compare} gives the density profile for $N=8192$
particles with an average density $\arho=0.05$ under the nonuniform field given
by \eqref{eq_wedge2} with $E_0=4$ and $A_0=0.1$. This profile shows the
condensation of a liquid-like layer from the homogeneous vapor phase in the
strong field region. The high density layer of width $\approx 2-3\sigma$ is
followed by a sharp interface and then a distinct vapor phase. The width of the
liquid-like layer grows as the fraction of energetically costly interface
molecules is reduced in larger systems. This effect is shown in the dash-dot and
solid
curves in Fig. \ref{fig_compare} where we increase the number of particles while
keeping the average density the same. Here, since we scale the simulation box
to keep the average density constant we also adjust the field in
\eqref{eq_wedge2} through $A_0$ such
that $E(z=D)$ is kept constant. In addition, the liquid-like
layer density also increases due to the decreased energetic cost of the
interface.

We explain this condensation by the fact that the nonuniform field is large only
in the vicinity of $z=0$. Hence, particles are
drawn to this region where they gain energetically both by aligning in the
stronger field and also from the LJ interaction which compensates for the loss
in entropy. The attractive short range part in the interaction is important
for the formation of a dense liquid layer. To show this we
also performed a simulation where we replace the LJ part of the interaction by
the purely repulsive WCA potential. The result for the dipolar WCA fluid is
shown in the dotted curve of \ref{fig_compare}; clearly, only a moderate
increase in the fluid density occurs and it follows the gradual decay of the
field.

We use the Stockmayer potential parameters for water
\cite{vanleeuwen1994}, which gives $\mu=1.56$ for the fluid and the
field magnitude of $E_0=4$ corresponds to a maximal local field of
$4.1$ V/nm. At least for water confined at the molecular scale such a
field is not unusual \cite{bratko2007}. Nonetheless, this field
magnitude is still 5-10 times larger than the fields required to
induce condensation in the mean field treatment \cite{efips_jpcb}. The
high field is a consequence of the costly interfacial region in the
small system we simulate and is expected to be reduced in the
thermodynamic limit $N, D\rightarrow \infty$.

Fig. \ref{fig_nuf1} shows the density profiles obtained as a function
of the nonuniform field magnitude, $E_0$. As $E_0$ increases the
condensation occurs rapidly starting at $E_0\gtrsim2.5$ albeit
gradually as as our system is finite. The inset of Fig. \ref{fig_nuf1}
gives a comparison between the field of \eqref{eq_wedge2} and a
nonuniform field of the same functional form and magnitude but
\emph{perpendicular} to the slab walls. This type of field is
obtained in the zero curvature limit for a capacitor consisting of
concentric cylinders\cite{efips_jpcb}.

\begin{figure}[!tb] 
\begin{center}
 \includegraphics[width=3.3in]{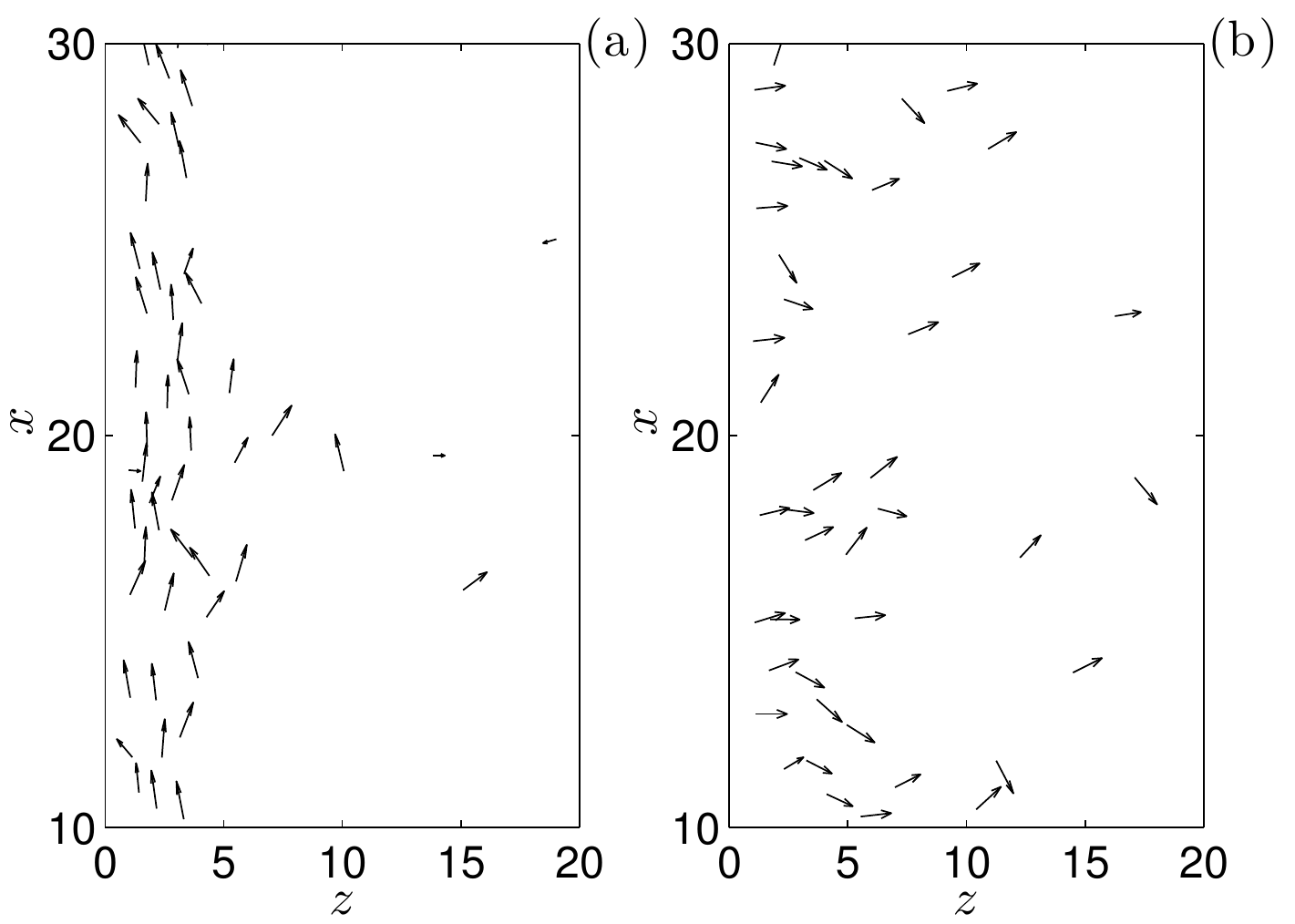}
 \caption{Snapshots of a thin cross section through the $y$ axis, focusing
on the strong field region. Arrows indicate the dipole moment's
$x$ and $z$ components scaled by a factor of 2. In (a) the electric field is
parallel to
the wall and $E_0=6$ while in (b) the field is perpendicular to
the wall and $E_0=12$ is larger. The maximal density near the wall is similar in
both panels but the interface is much wider in the perpendicular case due to
favored head to tail configurations.}
\label{fig_snap}
\end{center} 
\end{figure}

It is seen in Fig. \ref{fig_nuf1} that for
$E_0=4$ the density profile shows a clear condensate in the parallel case.
However, for the same value of $E_0$ in the 
perpendicular case, $\langle \rho(z) \rangle$ exhibits only a slight increase in
the fluid density near
the wall. Here, the field introduces a
competition between
alignment of dipoles parallel to the field, giving rise to the favored
head-to-tail configurations, and the creation of an interface parallel to the
field which disrupts these configurations \cite{tsori_rmp2009}. This is in
accord with the mean
field description in which the typical field required to induce condensation is
an order of magnitude larger in the cylindrical capacitor compared to the wedge
capacitor \cite{efips_jpcb}.
Only upon further increase of $E_0$ to large values of $E_0\gtrsim8$ a
significant increase in the density occurs. This is accompanied by large
oscillations of the density close to the wall in which the distance between
peaks is $\sigma$. This is typical when fields
perpendicular to the confining walls are applied to a high density dipolar
fluid \cite{lee1986}. The oscillatory domain is followed by an interfacial
region of width $\approx10\sigma$, which is large compared to the thin
interface of width $\approx4\sigma$ for parallel fields. Simulation
snapshots of a small segment of the system, shown in Fig. \ref{fig_snap},
illustrate how the orientational order leads to a wider interface in the
perpendicular case. We assume
that due to the large interfacial energetic penalty in perpendicular fields one
must simulate larger systems in order to observe clearly field induced
condensation in this case.

\begin{figure}[!tb] 
\begin{center}
 \includegraphics[width=3in]{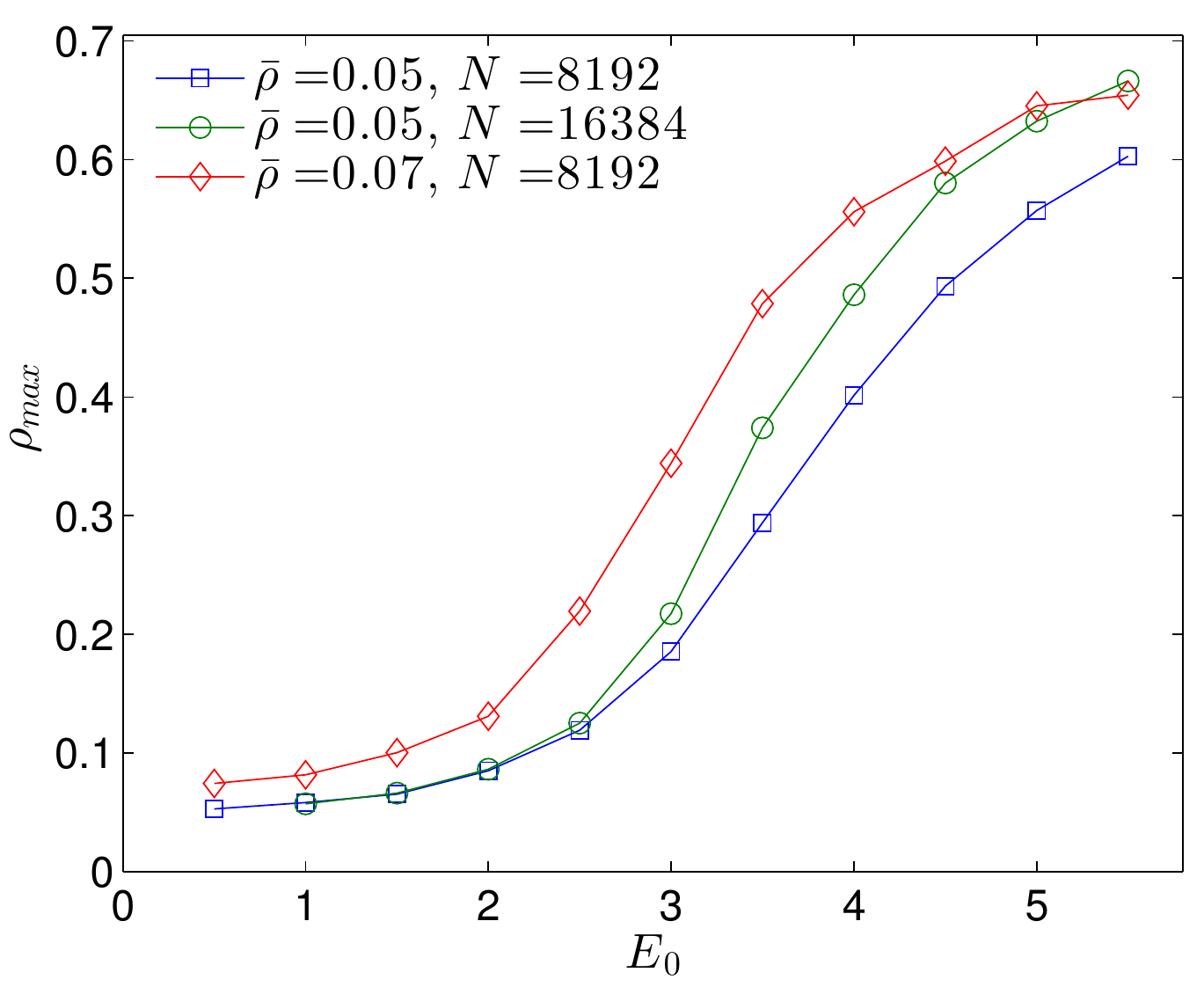}
 \caption{Maximal density as a function of the nonuniform field
  magnitude. $A_0=0.1$ in all curves while the average density or
  number of particles is varied.}
\label{fig_rho}
\end{center} 
\end{figure}

A hallmark of the first order transition in nonuniform fields
is a discontinuity in the surface density \cite{efips_jcp2}. The corresponding
quantity in the
simulation $\rho_{max}=\max(\langle \rho(z) \rangle)$ naturally occurs
close to the wall where the field is large. In Fig. \ref{fig_rho} we plot
$\rho_{max}$ as a function of the field magnitude. We find
that $\rho_{max}(E_0)$ has a sigmoid like shape, similar to the
mean field theory (see Fig. 10 in Ref. \cite{efips_jcp2}). However, since the
simulated system is finite $\rho_{max}(E_0)$ changes continuously.
Nonetheless, $\rho_{max}(E_0)$ grows more rapidly when the number of
particles is increased from $N=8192$ (squares)
to $N=16384$ (circles), suggesting that in the thermodynamic limit a first
order transition is realized.

Fig. \ref{fig_rho} also shows that increasing the average density to
$\arho=0.07$ (diamonds) results in a larger condensate density.
Although $\arho=0.07$ is inside the binodal for the bulk system, this
curve shows that one can utilize the nonuniform field to modify the
density profile in a dilute enough finite system, such as a colloidal
suspension.

\begin{figure}[!tb] 
\begin{center}
 \includegraphics[width=3in,clip]{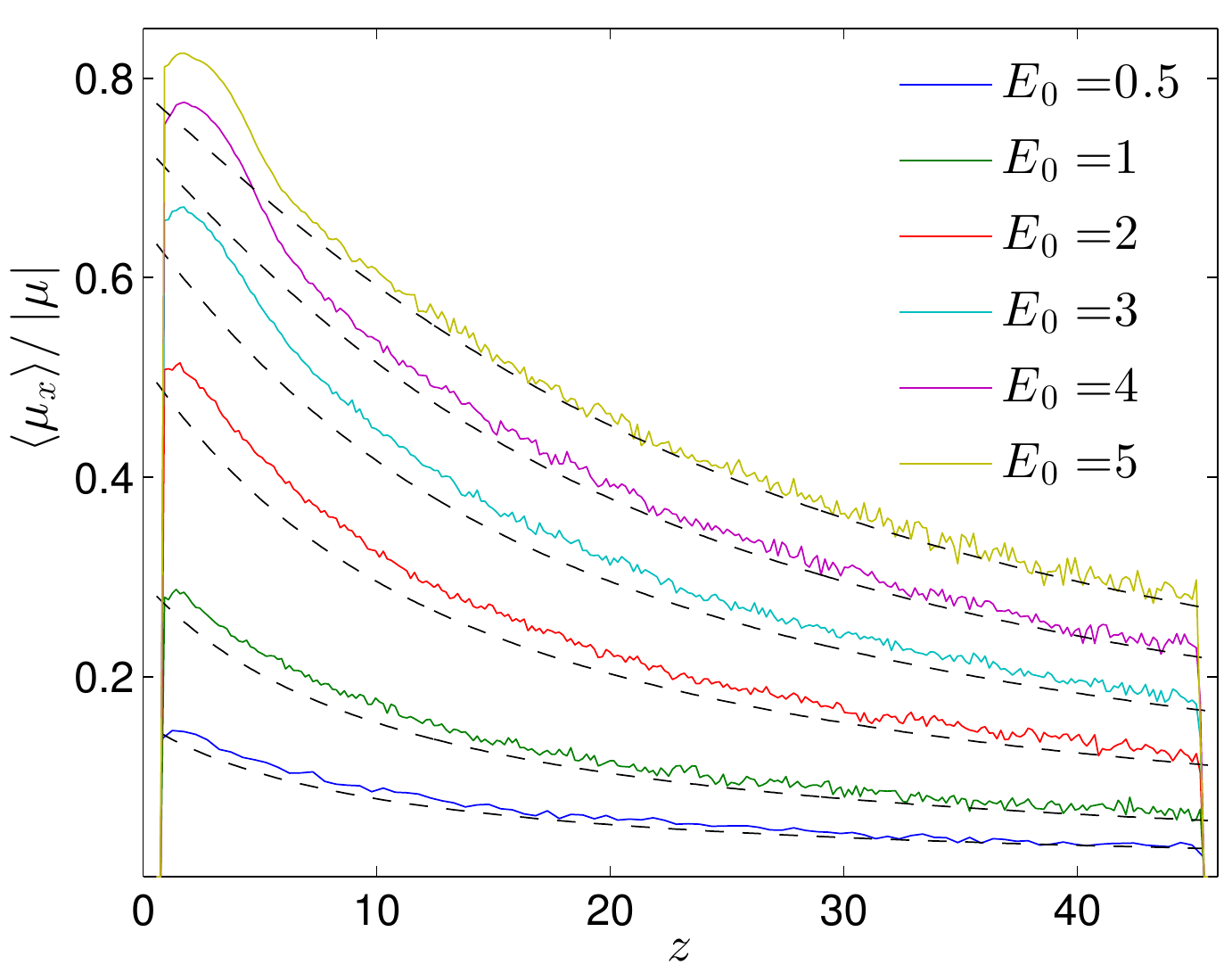}
 \caption{Solid curves: profiles of the scaled $x$ component of the
  dipole moment $\mu_x/\left|\mu\right|$ for several nonuniform field
  strengths. dashed curves: Debye theory prediction
  (\eqref{eq:debye}). }
\label{fig_mux}
\end{center} 
\end{figure}

Results for the average dipole moment in the direction of the field
$\langle \mu_x(z) \rangle$ are shown in the solid curves of Fig. \ref{fig_mux}.
These results are contrasted with the Debye theory \cite{debye_book} for an
ideal gas shown in the
dash-dot curves of Fig. \ref{fig_mux}. In the Debye theory:
\begin{equation}
 \label{eq:debye}
 \langle \mu_x(z) \rangle =\mu L(\alpha)
\end{equation}
where $L(\alpha)=\coth(\alpha)-1/\alpha$ is the Langevin function and
$\alpha=\mu \left| \bms{E}(z) \right| / T$. The simulation results
agree with the Debye theory in the dilute vapor region but deviate to
higher values in the dense liquid region. The deviation stems from the
oversimplified treatment of the dipoles orientation correlation in the
Debye theory \cite{bartke2006} as well as the unaccounted effect of
the short range LJ interaction on the orientational correlation.

\begin{figure}[!tb] 
\begin{center}
 \includegraphics[width=3in]{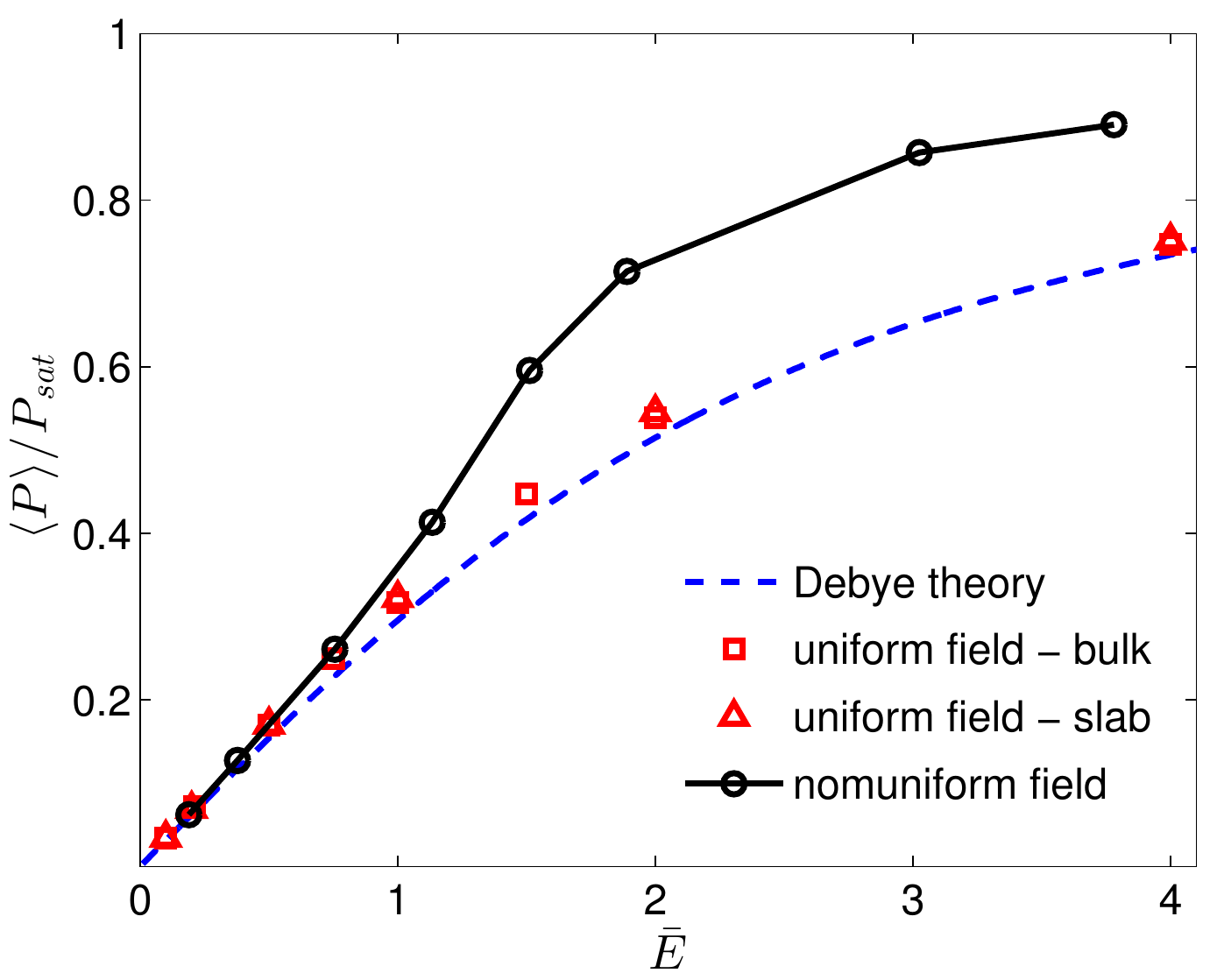}
 \caption{Scaled polarization $P/P_{sat}$ as a function of the average
  field $\bar{E}$. Bulk and slab systems in a uniform field exhibit a
  Langevin type polarization while in a nonuniform field the
  polarization is larger starting at $\bar{E}_0 \approx 1$.}
\label{fig_pol}
\end{center} 
\end{figure}

Further insight to the effect of the nonuniform field is gained by examining
the polarization of the system. $\langle \bms{P} \rangle =\langle \sum_i
\bms{\mu_i} \rangle$. Since in our case $E_z=E_y=0$, it follows from
\eqref{eq:debye} that for $E_x=const.$:
\begin{equation}
 \label{eq:pol}
 \langle P \rangle =P_{sat} L(\alpha)
\end{equation}
where $P_{sat}=N\mu$ is the saturation polarization of the system. We
compare the polarization for a uniform field in the bulk and in the
slab in Fig. \ref{fig_pol}. Simulation results in both cases are
almost identical and agreement with the Debye theory is very
good. This indicates that the bulk and slab system's response to a
uniform field is essentially the same here because we consider a large
enough slab, where the wall effects are negligible.

In order to compare results for a uniform field with those of a nonuniform
field we plot in the latter case the polarization as a function the
average field 
\begin{equation}
 \bar{E}_0=\frac{1}{D}\int_0^D E(z) {\rm d} z
\end{equation}
The solid curve in Fig. \ref{fig_pol} shows that the polarization for
the averaged nonuniform field is similar to that of the uniform field
up to $ \bar{E}_0 \approx 1$. This value corresponds to $E_0 \approx
2.5$ which in Fig. \ref{fig_nuf1} is where the fluid density near the
wall starts to increase. For $\bar{E}_0\gtrsim 1 $ the polarization
rapidly increases as the fluid condensates until it saturates at large
fields. Hence, the field induced condensation can be utilized to
amplify the electric response of a dilute dipolar system that will
otherwise follow the Langevin type response.

\section{conclusions} \label{sec:conc}

We studied the effect of a nonuniform field on a Stockmayer fluid via
molecular dynamics simulations. We find that a homogeneous vapor phase in the
canonical ensemble, unperturbed by a uniform field, undergoes a significant
structural change in a nonuniform field of the same magnitude. This
results in a sharp interface separating a liquid like region in the strong field
region and a dilute vapor where the field is weaker. We attribute this change to
the nonuniform field pulling the dipoles towards the strong field region
combined with the attractive short range part of the potential.

Our results indicate that a nonuniform field can be used to quite
sensitively control the density profile and hence the fluid properties
also in small closed systems. The mechanism we describe should be
applicable for a broad class of one-component systems, including
molecular fluids and colloidal suspensions. In fact, a nonuniform
field should promote phase separation in any dipolar system with an
inherent bistable nature
\cite{efips_jcp1,efips_jcp2,brunet2009,brunet2010}. We therefore
believe that the study of fluids in nonuniform fields merits further
experimental and theoretical attention.

\begin{acknowledgments}
We gratefully acknowledge numerous discussions with A. Arnold and O. Lenz. The
work has been performed under the HPC-EUROPA2 project (project number: 228398)
with the support of the European Commission - Capacities Area - Research
Infrastructures.
\end{acknowledgments}


\begin{thebibliography}{10}%
\makeatletter
\providecommand \@ifxundefined [1]{%
 \ifx #1\undefined \expandafter \@firstoftwo
 \else \expandafter \@secondoftwo
\fi
}%
\providecommand \@ifnum [1]{%
 \ifnum #1\expandafter \@firstoftwo
 \else \expandafter \@secondoftwo
\fi
}%
\providecommand \enquote [1]{``#1''}%
\providecommand \bibnamefont  [1]{#1}%
\providecommand \bibfnamefont [1]{#1}%
\providecommand \citenamefont [1]{#1}%
\providecommand\href[0]{\@sanitize\@href}%
\providecommand\@href[1]{\endgroup\@@startlink{#1}\endgroup\@@href}%
\providecommand\@@href[1]{#1\@@endlink}%
\providecommand \@sanitize [0]{\begingroup\catcode`\&12\catcode`\#12\relax}%
\@ifxundefined \pdfoutput {\@firstoftwo}{%
 \@ifnum{\z@=\pdfoutput}{\@firstoftwo}{\@secondoftwo}%
}{%
 \providecommand\@@startlink[1]{\leavevmode}%
 \providecommand\@@endlink[0]{}%
}{%
 \providecommand\@@startlink[1]{%
  \leavevmode
  \pdfstartlink
   attr{/Border[0 0 1 ]/H/I/C[0 1 1]}%
   user{/Subtype/Link/A<</Type/Action/S/URI/URI(#1)>>}%
  \relax
 }%
 \providecommand\@@endlink[0]{\pdfendlink}%
}%
\providecommand \url  [0]{\begingroup\@sanitize \@url }%
\providecommand \@url [1]{\endgroup\@href {#1}{\urlprefix}}%
\providecommand \urlprefix [0]{URL }%
\providecommand \Eprint[0]{\href }%
\@ifxundefined \urlstyle {%
  \providecommand \doi [1]{doi:\discretionary{}{}{}#1}%
}{%
  \providecommand \doi [0]{doi:\discretionary{}{}{}\begingroup
  \urlstyle{rm}\Url }%
}%
\providecommand \doibase [0]{http://dx.doi.org/}%
\providecommand \Doi[1]{\href{\doibase#1}}%
\providecommand \bibAnnote [3]{%
  \BibitemShut{#1}%
  \begin{quotation}\noindent
    \textsc{Key:}\ #2\\\textsc{Annotation:}\ #3%
  \end{quotation}%
}%
\providecommand \bibAnnoteFile [2]{%
  \IfFileExists{#2}{\bibAnnote {#1} {#2} {\input{#2}}}{}%
}%
\providecommand \typeout [0]{\immediate \write \m@ne }%
\providecommand \selectlanguage [0]{\@gobble}%
\providecommand \bibinfo [0]{\@secondoftwo}%
\providecommand \bibfield [0]{\@secondoftwo}%
\providecommand \translation [1]{[#1]}%
\providecommand \BibitemOpen[0]{}%
\providecommand \bibitemStop [0]{}%
\providecommand \bibitemNoStop [0]{.\EOS\space}%
\providecommand \EOS [0]{\spacefactor3000\relax}%
\providecommand \BibitemShut [1]{\csname bibitem#1\endcsname}%
\bibitem{klapp_review}%
  \BibitemOpen
  \bibfield{author}{%
  \bibinfo {author} {\bibfnamefont{S.~H.~L.}\ \bibnamefont{Klapp}},\ }%
  \bibfield{journal}{%
  \bibinfo {journal} {J. Phys.: Condens. Matter}\ }%
  \textbf{\bibinfo {volume} {17}},\ \bibinfo {pages} {R525} (\bibinfo {year}
  {2005})%
  \bibAnnoteFile{NoStop}{klapp_review}%
\bibitem{holm2005}%
  \BibitemOpen
  \bibfield{author}{%
  \bibinfo {author} {\bibfnamefont{C.}~\bibnamefont{Holm}}\ and\ \bibinfo
  {author} {\bibfnamefont{J.-J.}\ \bibnamefont{Weis}},\ }%
  \bibfield{journal}{%
  \bibinfo {journal} {Curr. Opin. Colloid Interface Sci.}\ }%
  \textbf{\bibinfo {volume} {10}},\ \bibinfo {pages} {133} (\bibinfo {year}
  {2005})%
  \bibAnnoteFile{NoStop}{holm2005}%
\bibitem{efdemix}%
  \BibitemOpen
  \bibfield{author}{%
  \bibinfo {author} {\bibfnamefont{Y.}~\bibnamefont{Tsori}}, \bibinfo {author}
  {\bibfnamefont{F.}~\bibnamefont{Tournilhac}},\ and\ \bibinfo {author}
  {\bibfnamefont{L.}~\bibnamefont{Leibler}},\ }%
  \bibfield{journal}{%
  \bibinfo {journal} {Nature}\ }%
  \textbf{\bibinfo {volume} {430}},\ \bibinfo {pages} {544} (\bibinfo {year}
  {2004})%
  \bibAnnoteFile{NoStop}{efdemix}%
\bibitem{efips_jpcb}%
  \BibitemOpen
  \bibfield{author}{%
  \bibinfo {author} {\bibfnamefont{S.}~\bibnamefont{Samin}}\ and\ \bibinfo
  {author} {\bibfnamefont{Y.}~\bibnamefont{Tsori}},\ }%
  \bibfield{journal}{%
  \bibinfo {journal} {J. Phys. Chem. B}\ }%
  \textbf{\bibinfo {volume} {115}},\ \bibinfo {pages} {75} (\bibinfo {year}
  {2011})%
  \bibAnnoteFile{NoStop}{efips_jpcb}%
\bibitem{efips_jcp1}%
  \BibitemOpen
  \bibfield{author}{%
  \bibinfo {author} {\bibfnamefont{G.}~\bibnamefont{Marcus}}, \bibinfo {author}
  {\bibfnamefont{S.}~\bibnamefont{Samin}},\ and\ \bibinfo {author}
  {\bibfnamefont{Y.}~\bibnamefont{Tsori}},\ }%
  \bibfield{journal}{%
  \Doi{10.1063/1.2965906}{\bibinfo {journal} {J. Chem. Phys.}}\ }%
  \textbf{\bibinfo {volume} {129}},\ \bibinfo {eid} {061101} (\bibinfo {year}
  {2008})%
  \bibAnnoteFile{NoStop}{efips_jcp1}%
\bibitem{efips_jcp2}%
  \BibitemOpen
  \bibfield{author}{%
  \bibinfo {author} {\bibfnamefont{S.}~\bibnamefont{Samin}}\ and\ \bibinfo
  {author} {\bibfnamefont{Y.}~\bibnamefont{Tsori}},\ }%
  \bibfield{journal}{%
  \Doi{10.1063/1.3257688}{\bibinfo {journal} {J. Chem. Phys.}}\ }%
  \textbf{\bibinfo {volume} {131}},\ \bibinfo {eid} {194102} (\bibinfo {year}
  {2009})%
  \bibAnnoteFile{NoStop}{efips_jcp2}%
\bibitem{chaikin_prl2006}%
  \BibitemOpen
  \bibfield{author}{%
  \bibinfo {author} {\bibfnamefont{M.~T.}\ \bibnamefont{Sullivan}}, \bibinfo
  {author} {\bibfnamefont{K.}~\bibnamefont{Zhao}}, \bibinfo {author}
  {\bibfnamefont{A.~D.}\ \bibnamefont{Hollingsworth}}, \bibinfo {author}
  {\bibfnamefont{R.~H.}\ \bibnamefont{Austin}}, \bibinfo {author}
  {\bibfnamefont{W.~B.}\ \bibnamefont{Russel}},\ and\ \bibinfo {author}
  {\bibfnamefont{P.~M.}\ \bibnamefont{Chaikin}},\ }%
  \bibfield{journal}{%
  \bibinfo {journal} {Phys. Rev. Lett.}\ }%
  \textbf{\bibinfo {volume} {96}},\ \bibinfo {pages} {015703} (\bibinfo {year}
  {2006})%
  \bibAnnoteFile{NoStop}{chaikin_prl2006}%
\bibitem{chaikin_jcp2008}%
  \BibitemOpen
  \bibfield{author}{%
  \bibinfo {author} {\bibfnamefont{M.~E.}\ \bibnamefont{Leunissen}}, \bibinfo
  {author} {\bibfnamefont{M.~T.}\ \bibnamefont{Sullivan}}, \bibinfo {author}
  {\bibfnamefont{P.~M.}\ \bibnamefont{Chaikin}},\ and\ \bibinfo {author}
  {\bibfnamefont{A.}~\bibnamefont{van Blaaderen}},\ }%
  \bibfield{journal}{%
  \bibinfo {journal} {J. Chem. Phys.}\ }%
  \textbf{\bibinfo {volume} {128}},\ \bibinfo {pages} {164508} (\bibinfo {year}
  {2008})%
  \bibAnnoteFile{NoStop}{chaikin_jcp2008}%
\bibitem{lumsdon2004}%
  \BibitemOpen
  \bibfield{author}{%
  \bibinfo {author} {\bibfnamefont{S.~O.}\ \bibnamefont{Lumsdon}}, \bibinfo
  {author} {\bibfnamefont{E.~W.}\ \bibnamefont{Kaler}},\ and\ \bibinfo {author}
  {\bibfnamefont{O.~D.}\ \bibnamefont{Velev}},\ }%
  \bibfield{journal}{%
  \Doi{10.1021/la035812y}{\bibinfo {journal} {Langmuir}}\ }%
  \textbf{\bibinfo {volume} {20}},\ \bibinfo {pages} {2108} (\bibinfo {year}
  {2004}),\ \bibinfo {note} {pMID: 15835659}%
  \bibAnnoteFile{NoStop}{lumsdon2004}%
\bibitem{bratko2007}%
  \BibitemOpen
  \bibfield{author}{%
  \bibinfo {author} {\bibfnamefont{D.}~\bibnamefont{Bratko}}, \bibinfo {author}
  {\bibfnamefont{C.~D.}\ \bibnamefont{Daub}}, \bibinfo {author}
  {\bibfnamefont{K.}~\bibnamefont{Leung}},\ and\ \bibinfo {author}
  {\bibfnamefont{A.}~\bibnamefont{Luzar}},\ }%
  \bibfield{journal}{%
  \Doi{10.1021/ja0659370}{\bibinfo {journal} {J. Am. Chem. Soc.}}\ }%
  \textbf{\bibinfo {volume} {129}},\ \bibinfo {pages} {2504} (\bibinfo {year}
  {2007})%
  \bibAnnoteFile{NoStop}{bratko2007}%
\bibitem{bratko2008}%
  \BibitemOpen
  \bibfield{author}{%
  \bibinfo {author} {\bibfnamefont{D.}~\bibnamefont{Bratko}}, \bibinfo {author}
  {\bibfnamefont{C.~D.}\ \bibnamefont{Daub}},\ and\ \bibinfo {author}
  {\bibfnamefont{A.}~\bibnamefont{Luzar}},\ }%
  \bibfield{journal}{%
  \Doi{10.1039/B809072F}{\bibinfo {journal} {Phys. Chem. Chem. Phys.}}\ }%
  \textbf{\bibinfo {volume} {10}},\ \bibinfo {pages} {6807} (\bibinfo {year}
  {2008})%
  \bibAnnoteFile{NoStop}{bratko2008}%
\bibitem{brunet2009}%
  \BibitemOpen
  \bibfield{author}{%
  \bibinfo {author} {\bibfnamefont{C.}~\bibnamefont{Brunet}}, \bibinfo {author}
  {\bibfnamefont{J.~G.}\ \bibnamefont{Malherbe}},\ and\ \bibinfo {author}
  {\bibfnamefont{S.}~\bibnamefont{Amokrane}},\ }%
  \bibfield{journal}{%
  \Doi{10.1063/1.3273870}{\bibinfo {journal} {J. Chem. Phys.}}\ }%
  \textbf{\bibinfo {volume} {131}},\ \bibinfo {eid} {221103} (\bibinfo {year}
  {2009})%
  \bibAnnoteFile{NoStop}{brunet2009}%
\bibitem{brunet2010}%
  \BibitemOpen
  \bibfield{author}{%
  \bibinfo {author} {\bibfnamefont{C.}~\bibnamefont{Brunet}}, \bibinfo {author}
  {\bibfnamefont{J.~G.}\ \bibnamefont{Malherbe}},\ and\ \bibinfo {author}
  {\bibfnamefont{S.}~\bibnamefont{Amokrane}},\ }%
  \bibfield{journal}{%
  \Doi{10.1103/PhysRevE.82.021504}{\bibinfo {journal} {Phys. Rev. E}}\ }%
  \textbf{\bibinfo {volume} {82}},\ \bibinfo {pages} {021504} (\bibinfo {year}
  {2010})%
  \bibAnnoteFile{NoStop}{brunet2010}%
\bibitem{gibbs_ensemble}%
  \BibitemOpen
  \bibfield{author}{%
  \bibinfo {author} {\bibfnamefont{A.~Z.}\ \bibnamefont{Panagiotopoulos}},\ }%
  \bibfield{journal}{%
  \bibinfo {journal} {Mol. Phys.}\ }%
  \textbf{\bibinfo {volume} {61}},\ \bibinfo {pages} {813} (\bibinfo {year}
  {1987})%
  \bibAnnoteFile{NoStop}{gibbs_ensemble}%
\bibitem{frenkel_book}%
  \BibitemOpen
  \bibfield{author}{%
  \bibinfo {author} {\bibfnamefont{D.}~\bibnamefont{Frenkel}}\ and\ \bibinfo
  {author} {\bibfnamefont{B.}~\bibnamefont{Smit}},\ }%
  \emph{\bibinfo {title} {Understanding Molecular Simulation}},\ \bibinfo
  {edition} {2nd}\ ed.\ (\bibinfo {publisher} {Academic Press},\ \bibinfo
  {address} {San Diego},\ \bibinfo {year} {2002})%
  \bibAnnoteFile{NoStop}{frenkel_book}%
\bibitem{Duane1987}%
  \BibitemOpen
  \bibfield{author}{%
  \bibinfo {author} {\bibfnamefont{S.}~\bibnamefont{Duane}}, \bibinfo {author}
  {\bibfnamefont{A.}~\bibnamefont{Kennedy}}, \bibinfo {author}
  {\bibfnamefont{B.~J.}\ \bibnamefont{Pendleton}},\ and\ \bibinfo {author}
  {\bibfnamefont{D.}~\bibnamefont{Roweth}},\ }%
  \bibfield{journal}{%
  \Doi{10.1016/0370-2693(87)91197-X}{\bibinfo {journal} {Phys. Lett. B}}\ }%
  \textbf{\bibinfo {volume} {195}},\ \bibinfo {pages} {216 } (\bibinfo {year}
  {1987}),\ ISSN \bibinfo {issn} {0370-2693}%
  \bibAnnoteFile{NoStop}{Duane1987}%
\bibitem{Mehlig1992}%
  \BibitemOpen
  \bibfield{author}{%
  \bibinfo {author} {\bibfnamefont{B.}~\bibnamefont{Mehlig}}, \bibinfo {author}
  {\bibfnamefont{D.~W.}\ \bibnamefont{Heermann}},\ and\ \bibinfo {author}
  {\bibfnamefont{B.~M.}\ \bibnamefont{Forrest}},\ }%
  \bibfield{journal}{%
  \Doi{10.1103/PhysRevB.45.679}{\bibinfo {journal} {Phys. Rev. B}}\ }%
  \textbf{\bibinfo {volume} {45}},\ \bibinfo {pages} {679} (\bibinfo {month}
  {Jan}\ \bibinfo {year} {1992})%
  \bibAnnoteFile{NoStop}{Mehlig1992}%
\bibitem{espresso}%
  \BibitemOpen
  \bibfield{author}{%
  \bibinfo {author} {\bibfnamefont{H.-J.}\ \bibnamefont{Limbach}}, \bibinfo
  {author} {\bibfnamefont{A.}~\bibnamefont{Arnold}}, \bibinfo {author}
  {\bibfnamefont{B.~A.}\ \bibnamefont{Mann}},\ and\ \bibinfo {author}
  {\bibfnamefont{C.}~\bibnamefont{Holm}},\ }%
  \bibfield{journal}{%
  \Doi{10.1016/j.cpc.2005.10.005}{\bibinfo {journal} {Comput. Phys. Commun.}}\
  }%
  \textbf{\bibinfo {volume} {174}},\ \bibinfo {pages} {704} (\bibinfo {year}
  {2006})%
  \bibAnnoteFile{NoStop}{espresso}%
\bibitem{desgranges2009}%
  \BibitemOpen
  \bibfield{author}{%
  \bibinfo {author} {\bibfnamefont{C.}~\bibnamefont{Desgranges}}\ and\ \bibinfo
  {author} {\bibfnamefont{J.}~\bibnamefont{Delhommelle}},\ }%
  \bibfield{journal}{%
  \Doi{10.1063/1.3158605}{\bibinfo {journal} {J. Chem. Phys.}}\ }%
  \textbf{\bibinfo {volume} {130}},\ \bibinfo {eid} {244109} (\bibinfo {year}
  {2009})%
  \bibAnnoteFile{NoStop}{desgranges2009}%
\bibitem{cerda2008}%
  \BibitemOpen
  \bibfield{author}{%
  \bibinfo {author} {\bibfnamefont{J.~J.}\ \bibnamefont{Cerda}}, \bibinfo
  {author} {\bibfnamefont{V.}~\bibnamefont{Ballenegger}}, \bibinfo {author}
  {\bibfnamefont{O.}~\bibnamefont{Lenz}},\ and\ \bibinfo {author}
  {\bibfnamefont{C.}~\bibnamefont{Holm}},\ }%
  \bibfield{journal}{%
  \Doi{10.1063/1.3000389}{\bibinfo {journal} {J. Chem. Phys.}}\ }%
  \textbf{\bibinfo {volume} {129}},\ \bibinfo {eid} {234104} (\bibinfo {year}
  {2008})%
  \bibAnnoteFile{NoStop}{cerda2008}%
\bibitem{brodka2004}%
  \BibitemOpen
  \bibfield{author}{%
  \bibinfo {author} {\bibfnamefont{A.}~\bibnamefont{Brodka}},\ }%
  \bibfield{journal}{%
  \bibinfo {journal} {Chem. Phys. Lett.}\ }%
  \textbf{\bibinfo {volume} {400}},\ \bibinfo {pages} {62} (\bibinfo {year}
  {2004})%
  \bibAnnoteFile{NoStop}{brodka2004}%
\bibitem{leeuwen1993}%
  \BibitemOpen
  \bibfield{author}{%
  \bibinfo {author} {\bibfnamefont{M.}~\bibnamefont{Van~Leeuwen}}, \bibinfo
  {author} {\bibfnamefont{B.}~\bibnamefont{Smit}},\ and\ \bibinfo {author}
  {\bibfnamefont{E.}~\bibnamefont{Hendriks}},\ }%
  \bibfield{journal}{%
  \bibinfo {journal} {Mol. Phys.}\ }%
  \textbf{\bibinfo {volume} {78}},\ \bibinfo {pages} {271} (\bibinfo {year}
  {1993})%
  \bibAnnoteFile{NoStop}{leeuwen1993}%
\bibitem{kiyohara1997}%
  \BibitemOpen
  \bibfield{author}{%
  \bibinfo {author} {\bibfnamefont{K.}~\bibnamefont{Kiyohara}}, \bibinfo
  {author} {\bibfnamefont{K.~E.}\ \bibnamefont{Gubbins}},\ and\ \bibinfo
  {author} {\bibfnamefont{A.~Z.}\ \bibnamefont{Panagiotopoulos}},\ }%
  \bibfield{journal}{%
  \bibinfo {journal} {J. Chem. Phys.}\ }%
  \textbf{\bibinfo {volume} {106}},\ \bibinfo {pages} {3338} (\bibinfo {year}
  {1997})%
  \bibAnnoteFile{NoStop}{kiyohara1997}%
\bibitem{stevens1995}%
  \BibitemOpen
  \bibfield{author}{%
  \bibinfo {author} {\bibfnamefont{M.~J.}\ \bibnamefont{Stevens}}\ and\
  \bibinfo {author} {\bibfnamefont{G.~S.}\ \bibnamefont{Grest}},\ }%
  \bibfield{journal}{%
  \bibinfo {journal} {Phys. Rev. E}\ }%
  \textbf{\bibinfo {volume} {51}},\ \bibinfo {pages} {5976} (\bibinfo {year}
  {1995})%
  \bibAnnoteFile{NoStop}{stevens1995}%
\bibitem{boda1996}%
  \BibitemOpen
  \bibfield{author}{%
  \bibinfo {author} {\bibfnamefont{D.}~\bibnamefont{Boda}}, \bibinfo {author}
  {\bibfnamefont{J.}~\bibnamefont{Winkleman}}, \bibinfo {author}
  {\bibfnamefont{J.}~\bibnamefont{Liszi}},\ and\ \bibinfo {author}
  {\bibfnamefont{I.}~\bibnamefont{Szalai}},\ }%
  \bibfield{journal}{%
  \bibinfo {journal} {Mol. Phys.}\ }%
  \textbf{\bibinfo {volume} {87}},\ \bibinfo {pages} {601} (\bibinfo {year}
  {1996})%
  \bibAnnoteFile{NoStop}{boda1996}%
\bibitem{kiyohara1999}%
  \BibitemOpen
  \bibfield{author}{%
  \bibinfo {author} {\bibfnamefont{K.}~\bibnamefont{Kiyohara}}, \bibinfo
  {author} {\bibfnamefont{K.~J.}\ \bibnamefont{Oh}}, \bibinfo {author}
  {\bibfnamefont{X.~C.}\ \bibnamefont{Zeng}},\ and\ \bibinfo {author}
  {\bibfnamefont{K.}~\bibnamefont{Ohta}},\ }%
  \bibfield{journal}{%
  \bibinfo {journal} {Mol. Simul.}\ }%
  \textbf{\bibinfo {volume} {23}},\ \bibinfo {pages} {95} (\bibinfo {year}
  {1999})%
  \bibAnnoteFile{NoStop}{kiyohara1999}%
\bibitem{szalai2003}%
  \BibitemOpen
  \bibfield{author}{%
  \bibinfo {author} {\bibfnamefont{I.}~\bibnamefont{Szalai}}, \bibinfo {author}
  {\bibfnamefont{K.-Y.}\ \bibnamefont{Chan}},\ and\ \bibinfo {author}
  {\bibfnamefont{Y.~W.}\ \bibnamefont{Tang}},\ }%
  \bibfield{journal}{%
  \bibinfo {journal} {Mol. Phys.}\ }%
  \textbf{\bibinfo {volume} {101}},\ \bibinfo {pages} {1819} (\bibinfo {year}
  {2003})%
  \bibAnnoteFile{NoStop}{szalai2003}%
\bibitem{jia2009}%
  \BibitemOpen
  \bibfield{author}{%
  \bibinfo {author} {\bibfnamefont{R.}~\bibnamefont{Jia}}\ and\ \bibinfo
  {author} {\bibfnamefont{R.}~\bibnamefont{Hentschke}},\ }%
  \bibfield{journal}{%
  \Doi{10.1103/PhysRevE.80.051502}{\bibinfo {journal} {Phys. Rev. E}}\ }%
  \textbf{\bibinfo {volume} {80}},\ \bibinfo {pages} {051502} (\bibinfo {year}
  {2009})%
  \bibAnnoteFile{NoStop}{jia2009}%
\bibitem{landau}%
  \BibitemOpen
  \bibfield{author}{%
  \bibinfo {author} {\bibfnamefont{L.~D.}\ \bibnamefont{Landau}}\ and\ \bibinfo
  {author} {\bibfnamefont{E.~M.}\ \bibnamefont{Lifshitz}},\ }%
  \emph{\bibinfo {title} {Elektrodinamika Sploshnykh Sred Chap. II, Sec. 18,
  Problem 1}}\ (\bibinfo {publisher} {Nauka},\ \bibinfo {address} {Moscow},\
  \bibinfo {year} {1957})%
  \bibAnnoteFile{NoStop}{landau}%
\bibitem{vanleeuwen1994}%
  \BibitemOpen
  \bibfield{author}{%
  \bibinfo {author} {\bibfnamefont{M.}~\bibnamefont{van Leeuwen}},\ }%
  \bibfield{journal}{%
  \bibinfo {journal} {Fluid Phase Equilib.}\ }%
  \textbf{\bibinfo {volume} {99}},\ \bibinfo {pages} {1 } (\bibinfo {year}
  {1994})%
  \bibAnnoteFile{NoStop}{vanleeuwen1994}%
\bibitem{bartke2007}%
  \BibitemOpen
  \bibfield{author}{%
  \bibinfo {author} {\bibfnamefont{J.}~\bibnamefont{Bartke}}\ and\ \bibinfo
  {author} {\bibfnamefont{R.}~\bibnamefont{Hentschke}},\ }%
  \bibfield{journal}{%
  \Doi{10.1103/PhysRevE.75.061503}{\bibinfo {journal} {Phys. Rev. E}}\ }%
  \textbf{\bibinfo {volume} {75}},\ \bibinfo {pages} {061503} (\bibinfo {year}
  {2007})%
  \bibAnnoteFile{NoStop}{bartke2007}%
\bibitem{richardi2008}%
  \BibitemOpen
  \bibfield{author}{%
  \bibinfo {author} {\bibfnamefont{J.}~\bibnamefont{Richardi}}, \bibinfo
  {author} {\bibfnamefont{M.~P.}\ \bibnamefont{Pileni}},\ and\ \bibinfo
  {author} {\bibfnamefont{J.-J.}\ \bibnamefont{Weis}},\ }%
  \bibfield{journal}{%
  \bibinfo {journal} {Phys. Rev. E}\ }%
  \textbf{\bibinfo {volume} {77}},\ \bibinfo {pages} {061510} (\bibinfo {year}
  {2008})%
  \bibAnnoteFile{NoStop}{richardi2008}%
\bibitem{tsori_rmp2009}%
  \BibitemOpen
  \bibfield{author}{%
  \bibinfo {author} {\bibfnamefont{Y.}~\bibnamefont{Tsori}},\ }%
  \bibfield{journal}{%
  \bibinfo {journal} {Rev. Mod. Phys.}\ }%
  \textbf{\bibinfo {volume} {81}},\ \bibinfo {pages} {1471} (\bibinfo {year}
  {2009})%
  \bibAnnoteFile{NoStop}{tsori_rmp2009}%
\bibitem{lee1986}%
  \BibitemOpen
  \bibfield{author}{%
  \bibinfo {author} {\bibfnamefont{S.~H.}\ \bibnamefont{Lee}}, \bibinfo
  {author} {\bibfnamefont{J.~C.}\ \bibnamefont{Rasaiah}},\ and\ \bibinfo
  {author} {\bibfnamefont{J.~B.}\ \bibnamefont{Hubbard}},\ }%
  \bibfield{journal}{%
  \Doi{10.1063/1.451663}{\bibinfo {journal} {J. Chem. Phys.}}\ }%
  \textbf{\bibinfo {volume} {85}},\ \bibinfo {pages} {5232} (\bibinfo {year}
  {1986})%
  \bibAnnoteFile{NoStop}{lee1986}%
\bibitem{debye_book}%
  \BibitemOpen
  \bibfield{author}{%
  \bibinfo {author} {\bibfnamefont{P.}~\bibnamefont{Debye}},\ }%
  \emph{\bibinfo {title} {Polar Molecules}}\ (\bibinfo {publisher} {Dover},\
  \bibinfo {address} {New York},\ \bibinfo {year} {1928})%
  \bibAnnoteFile{NoStop}{debye_book}%
\bibitem{bartke2006}%
  \BibitemOpen
  \bibfield{author}{%
  \bibinfo {author} {\bibfnamefont{J.}~\bibnamefont{Bartke}}\ and\ \bibinfo
  {author} {\bibfnamefont{R.}~\bibnamefont{Hentschke}},\ }%
  \bibfield{journal}{%
  \bibinfo {journal} {Mol. Phys.}\ }%
  \textbf{\bibinfo {volume} {104}},\ \bibinfo {pages} {3057} (\bibinfo {year}
  {2006})%
  \bibAnnoteFile{NoStop}{bartke2006}%
\end{thebibliography}
\end{document}